# Label-free image scanning microscopy for kHz super-resolution imaging and single particle tracking

**D**uc-**M**inh **T**a, **A**lberto **A**guilar, **and** **P**ierre **B**on*

*Xlim Research Institute, CNRS UMR 7252, Université de Limoges, Limoges, France.*
*\* pierre.bon@cnrs.fr*

**Abstract:** We report the modification of a label-free image scanning microscope (ISM) to perform asynchronous 2D imaging at 24kHz while keeping the lateral resolution gain and background rejection of a regular label-free ISM setup. Our method uses a resonant mirror oscillating at 12kHz for one-direction scanning and a chromatic line for instantaneous scanning in the other direction. We adapt optical photon reassignment in this scanning regime to perform fully optical super-resolution imaging. We exploit the kHz imaging capabilities of this confocal imaging system for single nanoparticle tracking down to 20nm for gold and 50nm for silica particles as well as imaging freely moving Lactobacillus with improved resolution.

## 1. Introduction

The ISM had been experimentally demonstrated in 2010[1] for fluorescence confocal microscopy, based on theoretical considerations by Sheppard in 1988[2]. The method allows to double the frequency support of a confocal microscope to perform super-resolution imaging. It consists of sampling the image produced by each point of a raster-scanning confocal microscope rather than just counting the total number of photons, as in classic confocal microscopy. The reconstruction process leads to resolution enhancement where each detected spot is reassigned at a double distance as compared to the actual scan in the sample. Image scanning microscopy is also named photon reassignment[3] or rescan confocal microscopy[4] based on different implementations. Even though it has been used mainly for fluorescence imaging, resolution enhancement can also be achieved for label-free imaging[5]. Particularly, for reflectance confocal microscopy, the reduction by $\sqrt{2}$ of the point spread function (PSF) conducts to a doubled frequency support as compared to regular confocal microscope[6]. It usually leads to resolution doubling even for quasi-periodic structures' imaging[7].

The original experimental approach of the ISM technique[1] requires numerical reassignment of all PSF-like images obtained on each point in the sample to get a super-resolution imaging effect. The whole process is slow and limits the acquisition rate. Therefore, to speed up the acquisition and simplify the data management, a fully optical photon reassignment can be implemented. The straightforward approach is to optically reassign each collected point on camera by a scanning mirror such as a galvanometer[4]. Then, to improve the scanning speed, a combination of galvanometer and resonant mirror has been proposed[6]. The fastest current versions, developed for fluorescence imaging, rely on parallelized point scanning using micro-lens arrays on spinning disk[8] or in a 2D periodic disposition[9]. 2D imaging at 100Hz is thus possible but with reduced background rejection compared to single-point scanning. The incoherent nature of fluorescence imaging reduces the choice for optical photon reassignment strategies; switching to label-free imaging unlocks different strategies.

In this paper, we describe a novel image-scanning microscopy scheme dedicated to label-free imaging with some applications. We demonstrate 2D super-resolution imaging at 24kHz without requiring any synchronization between the camera and the microscope optomechanical parts. Our setup is designed to work under the optical photon reassignment regime and is based on the combination of a double-sided resonant mirror[10] and a diffraction grating[11] to form a 2D image. After introducing our fast label-free ISM, we present the characterization of the system in terms of resolution and aberrations. The system is fast and sensitive enough to perform Single Particle Tracking (SPT) in water with nanoparticle size down to 20 nm for gold and 50 nm for silica. Finally, we illustrate the interest of super-resolution label-free imaging capability at high speed on small biological micro-organisms by capturing freely moving Lactobacillus in water.

## 2. Concept, setup and resolution

A chromatic line is generated in the (y) direction with a spatially coherent white source (super-continuum laser) and a diffraction grating. This approach, known as spectrally encoded confocal microscopy[11–13], allows us to perform instant imaging in one direction. In the label-free configuration, the chromatic line is coherent and elastically backscattered by the sample and can thus be recombined (using the same diffraction grating) into a single point; then it can be filtered by a unique pinhole (fig. 1) for a background rejection equivalent to single-point scanning confocal microscope. And, for fast scanning/de-scanning along the orthogonal (x) direction, a resonant mirror is used. Together, this strategy allows us to perform a 2D scanning at the resonant speed with equivalent de-scan and optimal background rejection capabilities.

Let us discuss how to perform asynchronous optical photon reassignment and achieve analogic super-resolution imaging. After background rejection by the pinhole, we use a second diffraction grating (DG2) with grooves density twice larger than the one in the illumination/descan path (DG1). Each chromatic (PSF) is separated twice as far along the (y) axis compared to the object plane. For the other (x) direction, the light beam is rescanned by the other face of the resonant mirror[10], but its diameter has been doubled as compared to the opposite face to achieve the doubled PSF separation along (x) in the detection plane. With this design, the image forms at the speed of the resonant mirror, and photons are reassigned optically for super-resolution. A complete super-resolved image is created at 24kHz, as our resonant mirror oscillates at 12kHz.

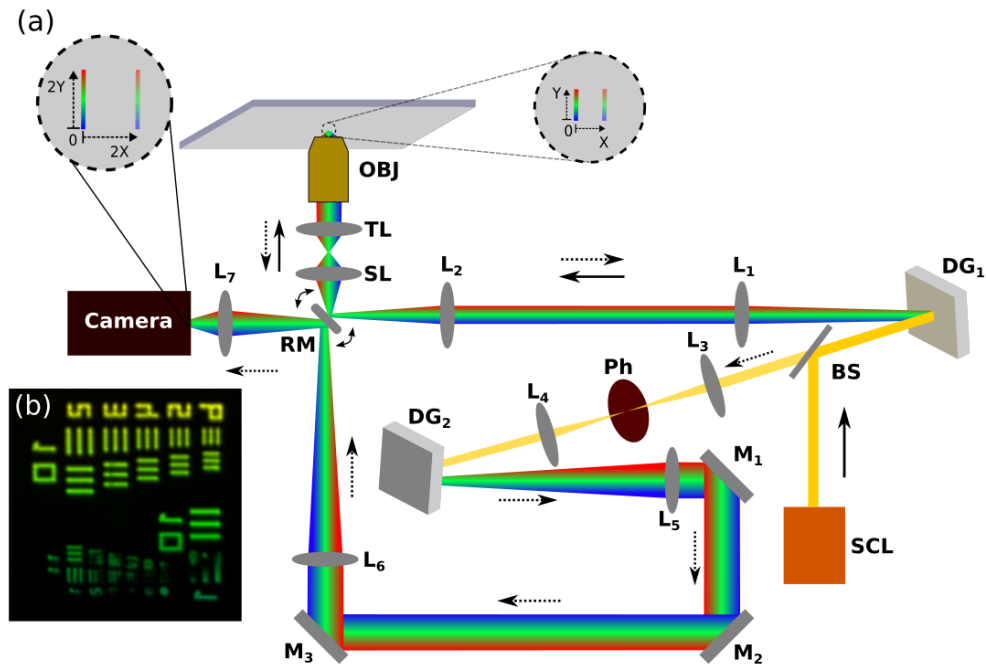

Figure 1: (a): Setup schematic. (SCL): Super Continuum Laser. (BS): Beam Splitter. ($DG_1$, $DG_2$): Diffraction Grating. ($L_1$-$L_7$): Lens. (RM): Resonant Mirror. (SL): Scan Lens. (TL): Tube Lens. (OBJ): Objective. (Ph): Pinhole. ($M_1$-$M_3$): Mirrors. (Camera) sCMOS Camera. (b): Example of the chromatic image formed in the image plane.

The detailed setup is shown in Fig. 1a. A supercontinuum laser beam (LEUKOS Electro-VIS 430) first goes to a 50:50 beam splitter plate (BS) and then to a blazed diffraction grating (DG1, 150 gr/mm, Edmund optics, 67% efficiency@500nm) to create a chromatic line using the first order of diffraction. Light is then adapted to the resonant mirror dimension by a 2:1 telescope (L1=300mm, L2=150mm). A double-sided resonant scan (RM, 12kHz, Novanta) is placed before the scan lens (SL, Thorlabs CLS-SL) and tube lens (TL, Thorlabs TL-400A). A water immersion objective (OBJ, Nikon 60X, NA=1.27 WI) focalized the light onto the sample. The back-scattered light from the sample follows the same path until it reaches DG1. There, the chromatic line is recombined into a white beam and is filtered by a 400μm pinhole inserted in a 4f system with L3=L4=200mm. Light is then diffracted by a second blazed diffraction grating (DG2, 300 gr/mm, Edmund optics, 67% efficiency@500nm); only the first order of diffraction is selected. Another 4f system (L5=L6=400mm) is used along with guiding mirrors to conjugate the pupil plane at the other side of the resonant mirror with a doubled diameter as compared to the opposite face. Finally, a camera (Orca Flash4 C11440, Hamamatsu) is placed at the focal plane of L7 to acquire the image. The 2D super-resolved image (displayed at 24 kHz) can be visualized directly without electronic synchronization, using a white screen instead of a camera (see Movie 1). The overall efficiency from the light backscattered by the sample up to the sensor plane is ≈ 20% (considering the efficiency of both descan and rescan gratings (DG1 and DG2) as well as the 50:50 beam splitter). Figure 1b shows what a chromatic image looks like.

The sample is placed on a motorized XY scanning stage (Thorlabs, MLS203), and the objective is mounted on a linear Z stage (Newport, M-426). The microscope is controlled via homemade LabVIEW software (NI). Manufacturer's software (HCImage) is used for image acquisition independently. Image processing and measurements have been performed with Labview, Matlab, and ImageJ.

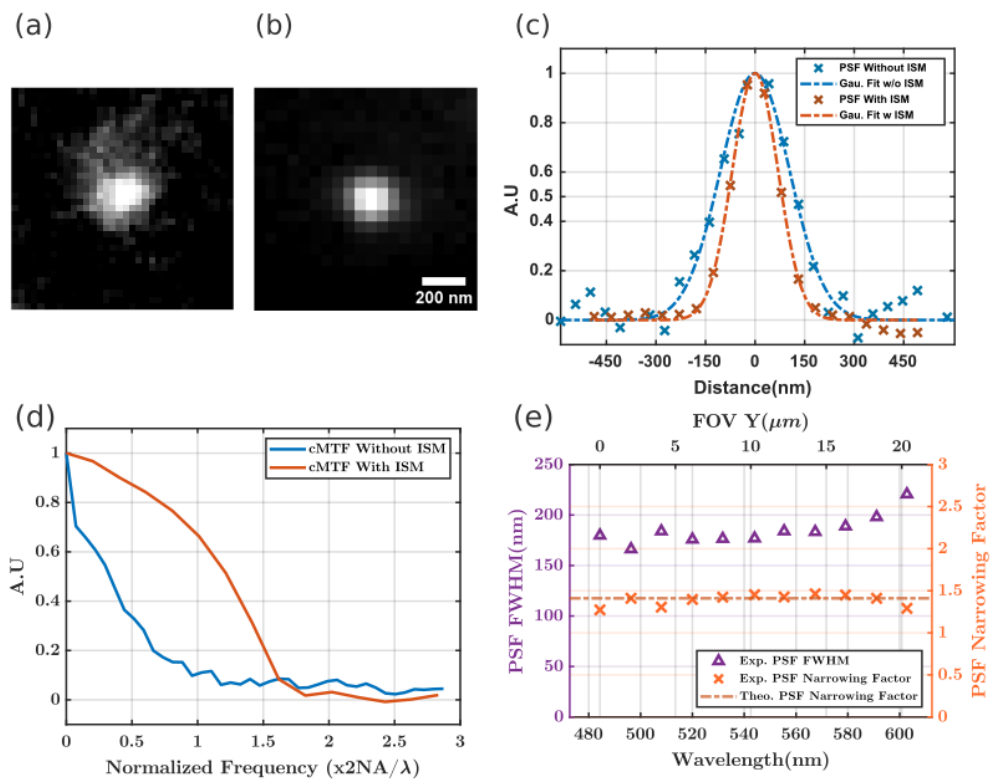

Figure 2: (a): Image of a 60nm Gold bead considered as the PSF for conventional confocal reflectance confocal microscope without ISM using a monodetector. (b): same as *(a)* with our super-resolved ISM. (c): Line profile of *(a), (b)* and its Gaussian fit for FWHM measurement. (d): coherent Modulation Transfer Function (cMTF) of the system without ISM and with ISM, frequency is normalized to theoretical diffraction limit frequency. (e): FWHM of PSFs at different wavelengths in Field of View (FOV) and PSF narrowing factor (expected to be $\sqrt{2}$ in dash-dotted line).

To characterize the resolution of the system, we image 60nm gold nanoparticles (Sigma Aldrich, 742015) immobilized on a #1.5 coverslip, covered by microscope immersion oil to minimize the reflection from the coverslip in the nanoparticle plane. A conventional reflectance confocal microscope without ISM has been built, and it shares the same objective for resolution comparison. For this microscope, a monochromatic 520nm laser is scanned via a 2D galvanometer mirror system (GVS002, Thorlabs) and injected between L1 et DG1 by a flip mirror (not represented in Fig.1). After de-scanning with the galvanometer mirrors, a pinhole of 1 Airy unit (AU) is used to reject the background of the backscattered light and a photodiode synchronized with the 2D scanner is used to reconstruct the confocal image.

Figure 2a shows the PSF image from this conventional reflectance confocal microscope, and Figure 2b is the PSF image obtained from our chromatic asynchronous ISM. The intensity profiles are plotted in Figure 1c and fitted by a Gaussian function to find each full width at half maximum (FWHM). These FWHM values are 240±20 nm for conventional confocal reflectance microscope and 170±10 nm for chromatic ISM @λ=500nm, NA=1.27. The PSF narrowing factor is measured at 1.4±0.1 close to the expected $\sqrt{2}$ gain factor for ISM methods[14]. The coherent modulation transfer function is calculated by taking the 2D Fourier Transform of each PSF and then average radially. The result is presented in Fig. 2d where the gain factor of frequency support could reach 1.8 (for a theoretical maximum value of 2). Due to the different wavelengths used for imaging, we also study the PSF sharpening dependency across the whole field of view (FOV). Fig. 2e shows the PSF narrowing factor compared to the theoretical PSF size across a FOV of 20μm, which corresponds to wavelength spread from 480nm to 600nm (so 540 nm ±11%).-The chromatic ISM PSF consistently outperforms the theoretical diffraction limited value by at least 1.3x.

### 3. Single particle tracking

The first study we performed with our asynchronous label-free ISM is single particle detection and tracking. SPT is crucial for characterizing nano-objects[15] because the Brownian motion of particles is directly related to the particle size. For non-actively moving particles, the smaller the faster. The mean squared displacement (MSD) of particles between successive frames can determine the size of each particle tracked, given the viscosity of the medium. If the particle is moving relatively slowly (MSD << PSF size), the localization error is determined by the total number collected photons as predicted by Fisher information theory and Cramer-Rao lower bound[16,17]. However, if the particle is quick, it can distort the acquired image, therefore increasing localization error. Moreover, since the particle size determination from its trajectory is converging in $\sqrt{M}$, with M the tracking length, it is important to track the particle during a large number of frames[18]. A fast imaging system is therefore interesting for single particle tracking to reduce the image distortion and increase the tracking length of a particle.

In super localization approaches *via* PSF fitting, the reduction of the PSF size is also an important aspect since the localization precision is directly proportional to the PSF diameter. Widefield structured illumination has been proposed to enhance the localization precision in fluorescence single particle tracking[19,20]. In fluorescence microscopy, the signal noise ratio (SNR) can be improved via background reduction using confocal laser scanning microscopy scheme[21,22].

Thanks to resolution enhancement, background rejection, and fast acquisition speed of our system, label-free single particle tracking is possible with improved SNR compared to wide-field imaging. In our case, the PSF size reduction by $\sqrt{2}$ (i) results in an increase of amplitude by 2 and (ii) unlocks tracking at higher density. As expected, the static localization precision in standard deviation $\sigma_x$ (by tracking multiple fixed 60 nm gold nanoparticles within the field of view to avoid thermomechanical drifts [23]) follows a linear dependence concerning to $\frac{1}{\sqrt{N_p}}$, with $N_p$ the number of detected photons (Fig. 3.a) and a quadratic evolution with the defocus (fig. 3.b). For $N_p = 3500$, and no defocus, the localization precision reaches 4.3 nm.

With our setup, we demonstrate the detectability and tracking of 20 nm gold nanoparticles (sigma Aldrich, 741965), 100 nm and 50 nm silica nanoparticles (Sigma Aldrich, 797936), in water, at a few tens of microns in the solution, using 8kW/cm² of average illumination power (25MW/cm² peak power). The acquisition rate of our camera has been fixed at 1000Hz (maximum possible), and the sampling has been optimized for SPT[16] at 96 nm/pixel in the object space. The PSF is sampled with 1.85 pixels, and a crop around a single particle in each acquisition along with its intensity profile is presented in the inset of Fig. 3.c for signal-to-noise ratio comparison. Although they are reflectance images, the super-localization and tracking algorithms developed for fluorescence imaging can be applied since the signal looks very similar (see Movie 2). The tracking process for each particle was done by the plugin Trackmate in ImageJ[24]. The MSD was calculated for each particle and linear fitted to find the diffusion coefficient. the hydrodynamic size is then deduced from the Stoke-Einstein equation[25]. Fig. 3.d shows good agreement between the measured distribution of the hydrodynamic size and manufacturer size which illustrate that the system can detect and track gold nanoparticle of size down to 20 nm and silica particle of size down to 50 nm at depth in the sample.

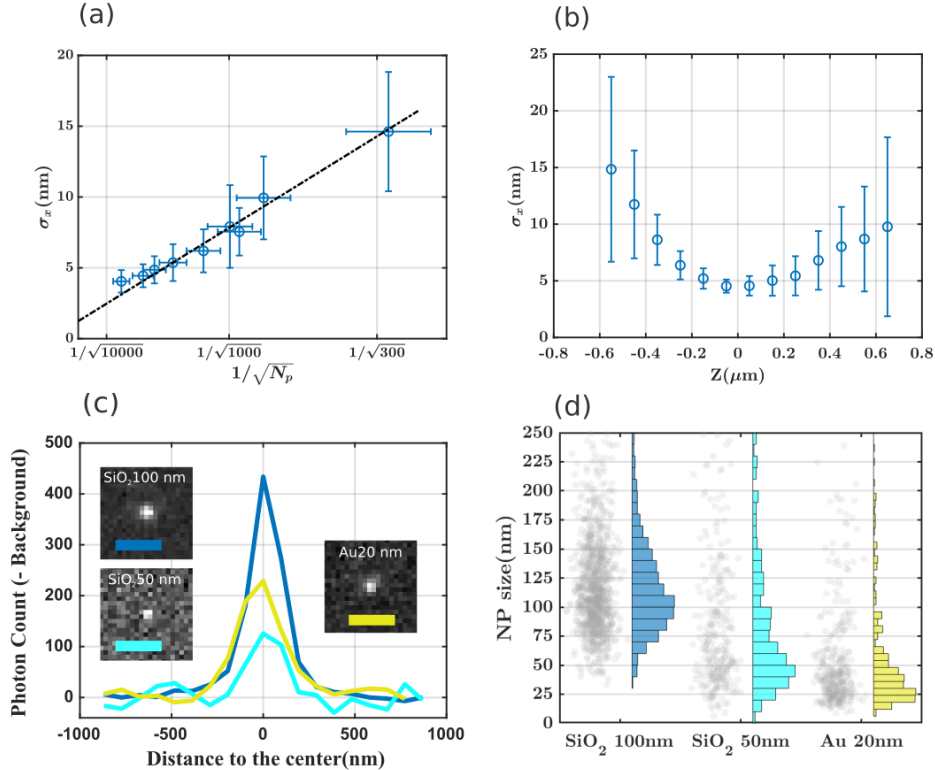

Figure 3: (a): Static localization error $\sigma_x$ as a function of the number of photons $N_p$. (b): Static localization error $\sigma_x$ at a different focus for $N_p = 3500$ photons. (c): Crop of single

nanoparticles acquired @1000Hz for SiO$_2$ 100nm, SiO$_2$ 50nm, and Au 20nm moving in the water and their intensity profile (with background subtracted), the scale bar is 1µm and color-coded according to the intensity profile. (f): Distribution of nanoparticle hydrodynamic size for each type of particle obtained from diffusion equation and Stoke-Einstein relation.

### 4. Imaging of freely moving bacteria

We then moved to biological sample label-free where fast and super-resolved imaging is important. We have considered Lactobacillus moving freely in water. These non-motile bacteria are commonly found in yogurt and cheeses and have a rod shape of dimension $\{1 - 1.5\} \times \{0.7 - 1\}$ µm [26]. Capturing freely-moving bacteria would require a high-speed microscope with good resolution. Typically, agarose immobilizes the bacteria on the coverslip surface to avoid bacteria fast movements. In our experiment, there was no immobilization and the sample consists in a small portion of the rehydrated probiotic, putted directly on the coverslip for observation. Images are acquired at 100Hz @ 200W/cm² average illumination power. These illuminations generate high SNR images and can be reduced more in power for long-duration imaging without compromising the measurements. Fig. 4a demonstrates an image of the sample where a dividing bacterium is followed easily during a few seconds at 20µm above the coverslip. Different instants when the bacteria are close to focus are also shown in Fig. 4b-e (see Movie 3 for full acquisition). The backscattered intensity can be interpreted as the local dry mass density, which is higher at the tips of bacteria where the protein complex responsible for division during cytokinesis is present[27]. This signal is proportional to the refractive index mismatch with the medium, far from absorption and optical resonance. In Figure 4f and Figure 4g, the measured diameter (600 nm) and length of the imaged bacteria during its mitosis ($2 \times 1.5$ µm) reach the tabulated values. These dimensions are close to conventional optical method resolution and our resolution improvement along with speed enhancement pave the way to the study of small micro-organisms in their native and unmodified environment.

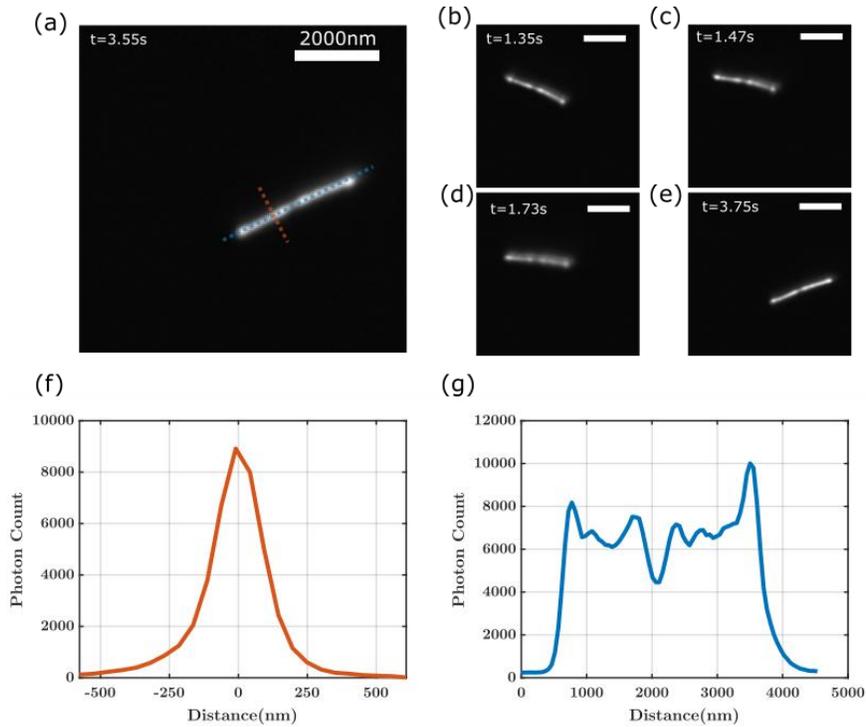

Figure 4: (a): Raw Image of Lactobacillus. (b-e): Image of Lactobacillus at different instants. (f), (g): Line profile along its length and cross section respectively.

**CONCLUSION**

We have introduced a new approach to label-free microscopy using ISM, enabling 2D super-resolution imaging at a speed of 24kHz, which is limited by the resonant mirror. The demonstration with nanoparticle tracking at 1000 frames per second, and imaging of moving Lactobacillus show the capability of our system in terms of speed, localization and resolution. The interest of super-localization with ISM lies both in an increased SNR (and so in localization precision[28]) and compatibility with higher density solution of particles. For live sample imaging, the label-free high-speed super-resolution imaging capability is key for small sample imaging (such as bacteria or viruses) as well as for transient phenomenon detection in tissues. Currently, the camera can capture 20% of scattered light from the objective, allowing for fast imaging with illumination compatible with live sample imaging. This could be further improved using prisms rather than diffraction gratings but with an increased risk of aberration and chromatic focal shifts.

For 3D imaging, we did not try to perform high-speed focusing. However, since our setup remains fully compatible with classical focusing methods, it is possible to scan faster in the Z direction by introducing an adaptive lens whose focal length can be varied electrically[29]. Moreover, the scanning rate could be further enhanced by implementing the technique outlined in a recent article, which involved using a lens array to multiply the scan speed.[30].

Our asynchronous super-resolution scheme avoids any synchronization issues and post-processing artifacts that could occur in structured illumination approaches or numerical ISM.

We see potential for applications beyond biology and nanoparticle analysis, including –fast control of surface nano-structuring[7].

Acknowledgments: This work was supported by CNRS and University of Limoges. This project has received funding from the European Research Council (ERC) under the European Union's Horizon 2020 research and innovation program (grant agreement No. [848645]).

1.	Müller, C. B. & Enderlein, J. Image Scanning Microscopy. *Phys. Rev. Lett.* **104**, 198101 (2010).

2.	Sheppard, C. J. R. Super-Resolution in Confocal Imaging.

3.	Roth, S., Sheppard, C. J., Wicker, K. & Heintzmann, R. Optical photon reassignment microscopy (OPRA). *Opt. Nanoscopy* **2**, 5 (2013).

4.	De Luca, G. M. R. *et al.* Re-scan confocal microscopy: scanning twice for better resolution. *Biomed. Opt. Express* **4**, 2644 (2013).

5.	Aguilar, A., Boyreau, A. & Bon, P. Label-free super-resolution imaging below 90-nm using photon-reassignment. *Open Res. Eur.* **1**, 3 (2021).

6.	DuBose, T. B., LaRocca, F., Farsiu, S. & Izatt, J. A. Super-resolution retinal imaging using optically reassigned scanning laser ophthalmoscopy. *Nat. Photonics* **13**, 257–262 (2019).

7.	Aguilar, A. *et al.* Nondestructive inspection of surface nanostructuring using label-free optical super-resolution imaging. (2023) doi:10/document.

8.	Azuma, T. & Kei, T. Super-resolution spinning-disk confocal microscopy using optical photon reassignment. *Opt. Express* **23**, 15003–15011 (2015).

9.	Hayashi, S. & Okada, Y. Ultrafast superresolution fluorescence imaging with spinning disk confocal microscope optics. *Mol. Biol. Cell* **26**, 1743–1751 (2015).

10.	Brakenhoff, G. J. & Visscher, K. Confocal imaging with bilateral scanning and array detectors. *J. Microsc.* **165**, 139–146 (1992).

11.	Tearney, G. J., Webb, R. H. & Bouma, B. E. Spectrally encoded confocal microscopy. *Opt. Lett.* **23**, 1152–1154 (1998).

12.	Boudoux, C. *et al.* Rapid wavelength-swept spectrally encoded confocal microscopy. *Opt. Express* **13**, 8214–8221 (2005).

13.	Kim, S. *et al.* Spectrally encoded slit confocal microscopy using a wavelength-swept laser. *J. Biomed. Opt.* **20**, 036016 (2015).

14.	Gregor, I. & Enderlein, J. Image scanning microscopy. *Curr. Opin. Chem. Biol.* **51**, 74–83 (2019).

15.	Nguyen, M.-C. *et al.* Label-free single nanoparticle identification and characterization including infectious emergent virus. Preprint at https://doi.org/10.48550/arXiv.2301.02542 (2023).

16.	Chao, J., Ward, E. S. & Ober, R. J. Fisher information theory for parameter estimation in single molecule microscopy: tutorial. *JOSA A* **33**, B36–B57 (2016).

17.	Wernet, M. P. & Pline, A. Particle displacement tracking technique and Cramer-Rao lower bound error in centroid estimates from CCD imagery. *Exp. Fluids* **15**, 295–307 (1993).

18.	Michalet, X. Mean square displacement analysis of single-particle trajectories with localization error: Brownian motion in an isotropic medium. *Phys. Rev. E* **82**, 041914 (2010).

19.	Cnossen, J. *et al.* Localization microscopy at doubled precision with patterned illumination. *Nat. Methods* **17**, 59–63 (2020).

20.	Jouchet, P., Poüs, C., Fort, E. & Lévêque-Fort, S. Time-modulated excitation for enhanced single-molecule localization microscopy. *Philos. Trans. R. Soc. Math. Phys. Eng. Sci.* **380**, 20200299 (2022).

21.	Vukojević, V. *et al.* Quantitative single-molecule imaging by confocal laser scanning microscopy. *Proc. Natl. Acad. Sci.* **105**, 18176–18181 (2008).

22.	Varela, J. A. *et al.* Targeting neurotransmitter receptors with nanoparticles in vivo allows single-molecule tracking in acute brain slices. *Nat Commun* **7**, (2016).

23.	Bon, P. *et al.* Three-dimensional nanometre localization of nanoparticles to enhance super-resolution microscopy. *Nat Commun* **6**, 7764- (2015).

24.	Tinevez, J.-Y. *et al.* TrackMate: An open and extensible platform for single-particle tracking. *Methods* **115**, 80–90 (2017).

25.	Pecora, R. Dynamic Light Scattering Measurement of Nanometer Particles in Liquids. *J. Nanoparticle Res.* **2**, 123–131 (2000).

26.	Schär-Zammaretti, P. & Ubbink, J. The Cell Wall of Lactic Acid Bacteria: Surface Constituents and Macromolecular Conformations. *Biophys. J.* **85**, 4076 (2003).

27.	Egan, A. J. F. & Vollmer, W. The physiology of bacterial cell division. *Ann. N. Y. Acad. Sci.* **1277**, 8–28 (2013).

28.	Ober, R. J., Ram, S. & Ward, E. S. Localization accuracy in single-molecule microscopy. *Biophys J* **86**, 1185–1200 (2004).

29.	Koukourakis, N. *et al.* Axial scanning in confocal microscopy employing adaptive lenses (CAL). *Opt. Express* **22**, 6025–6039 (2014).

30.	Xiao, S., Giblin, J. T., Boas, D. A. & Mertz, J. High-throughput deep tissue two-photon microscopy at kilohertz frame rates. *Optica* **10**, 763–769 (2023).